\documentclass[useAMS,usenatbib]{mn2e}
\usepackage{amsmath}
\usepackage{graphicx}
\usepackage{xcolor}
\numberwithin{equation}{section}
\usepackage{hhline}
\usepackage{cancel}
\usepackage[normalem]{ulem}

\newcommand{\be}{\begin{equation}}
\newcommand{\ee}{\end{equation}}

\newcommand{\e}{\times10^}
\newcommand{\Mmin}{M_{\rm{CO,min}}}

\newcommand{\co}{\rm{CO}}

\newcommand{\tco}{\left<T_{\co}\right>}

\newcommand{\fduty}{f_{\mathrm{duty}}}

\newcommand{\Nm}{\left<N\right>}
\newcommand{\vk}{\mathbf{k}}
\newcommand{\vx}{\mathbf{x}}
\newcommand{\tavg}{\left<T\right>}

\newcommand{\lya}{\rm{Ly}\alpha}

\begin{document}

\title[Masking line foregrounds in intensity mapping surveys]{Masking line foregrounds in intensity mapping surveys}
\author[Patrick C. Breysse, Ely D. Kovetz, and Marc Kamionkowski]{Patrick C. Breysse,$^{1}$\thanks{pbreysse@pha.jhu.edu (PCB); elykovetz@gmail.com (EDK); kamion@pha.jhu.edu (MK)} Ely D. Kovetz,$^{1}$\footnotemark[1] and Marc Kamionkowski$^{1}$\footnotemark[1] \\
$^{1}$ Department of Physics and Astronomy, Johns Hopkins University, Baltimore, MD 21218 USA}

\maketitle
\label{firstpage}

\begin{abstract}
We address the problem of line confusion in intensity mapping surveys and explore the possibility to mitigate  line foreground contamination by progressively masking the brightest pixels in the observed map. We consider experiments targeting CO(1-0) at $z=3$, $\lya$ at $z=7$, and CII at $z=7$, and use simulated intensity maps, which include both clustering and shot noise components of the signal and possible foregrounds, in order to test the efficiency of our method. We find that for CO and $\lya$ it is quite possible to remove most of the foreground contribution from the maps via only $1\%-3\%$ pixel masking. The CII maps will be more difficult to clean, however, due to instrumental constraints and the high-intensity foreground contamination involved. While the masking procedure sacrifices much of the astrophysical information present in our maps, we demonstrate that useful cosmological information in the targeted lines can be successfully retrieved.
\end{abstract}


\section{Introduction}
Intensity mapping is a powerful new technique for studying the large-scale structure of the universe.  By observing the large-scale fluctuations in the intensity of some chosen spectral line, it is possible to study a population of galaxies without needing the kind of depth and sensitivity required to find each galaxy individually \citep{mmr}.  The intensity of a line in a given volume depends both on the detailed astrophysical processes in the emission region as well as the underlying dark matter density.  Thus intensity mapping surveys contain a wealth of cosmological and astrophysical information which is difficult to obtain using other methods.

There are many lines that can be used for intensity mapping.  A number of experiments either proposed or in progress are seeking to study the 21 cm fine structure line in neutral hydrogen, such as the Square Kilometer Array (SKA) \citep{ska},  Low Frequency Array (LOFAR) \citep{lofar}, the Precision Array for Probing the Epoch or Reionization (PAPER) \citep{paper}, the Giant Meterwave Radio Telescope (GMRT) \citep{gmrt}, and the Murchison Widefield Array (MWA) \citep{mwa}.  We consider here three other commonly discussed lines:  the CO(1-0) rotational line, \citep{righi,lidz,pullen,BKK}, the $\lya$ hydrogen line \citep{pullenb,gonglya}, and the 157.7 $\mu$m CII fine structure line \citep{gcs,S14}.  Other lines which have been considered in the literature which we do not study here include molecular hydrogen \citep{H2} and HeII \citep{He2}.

Several experiments are planned to study these lines, including the Carbon MonOxide Mapping Array Pathfinder (COMAP) targeting CO \citep{li}, the Spectro-Photometer for the History of the Universe, Epoch of Reionization, and Ices Explorer (SPHEREx) targeting $\lya$ \citep{spherex}, and the Tomographic Ionized-Carbon Mapping Experiment (TIME) targeting CII \citep{TIME}. As we shall see, each of these lines has its own dependence on the conditions within the galaxies it is  emitted from, and studying each line has its own set of challenges. 

One major difficulty in intensity mapping surveys is the problem of foregrounds.  Every intensity mapping survey will have to deal with many types of foreground emission.  Foregrounds with continuum frequency spectra such as dust or synchrotron are problematic, but the removal of these foregrounds is a well studied problem, especially for 21 cm (see for example \citet{21fg1,21fg2}).  Line foregrounds are a more difficult problem.  If a spectral line other than the target is redshifted into the same observing band, it is not easy to tell the two lines apart.  21 cm surveys are not expected to suffer from this issue since there are so few lines at such low frequencies \citep{OH}. However surveys in other lines will require a better understanding of possible line foregrounds.  As we will show below, CO, $\lya$, and CII all have potential confusing lines.

One method which can be used to remove these foreground lines is to cross correlate an intensity map with another map in a different frequency, or with some other tracer of large-scale structure \citep{vl}.  Though each map will have its own foregrounds, the signals from the two maps will be correlated, and their foregrounds will not, leaving behind a cross spectrum which only depends on the two target populations.  However, this method has two issues:  The first is simply that it requires a more complicated (and costly) observation, since it requires observation of two signals.  Secondly, it is difficult to reconstruct the auto power spectrum of the target line from the cross spectrum of the two maps (\citet{gonglya}, hereafter G14).  Thus we seek a method to obtain the auto spectrum of a single map without foreground contamination.

Though ``foreground" lines could technically come from lower or higher redshifts than the target line, only lower redshift lines are likely to pose a problem, as the signal from an intensity mapping survey typically grows weaker with redshift.  The fact that these lines come from lower redshifts offers a potential way to remove them.  At lower redshifts, galaxy masses tend to be larger, so we expect there to be more very bright sources of a foreground line than a target line.  This means that the brightest pixels in a survey will tend to be foreground galaxies, and the foreground contamination can be at least partially removed by masking out the brightest pixels in a survey.  This technique was discussed by \citet{vtl}, who found that it tends to bias the target power spectrum because some signal is masked along with the foregrounds.  However, as we will show below, much of the \emph{cosmological} information in the power spectrum is preserved after masking, even though most of the \emph{astrophysical} information is lost.  G14 also explored this technique for $\lya$, though they do not appear to show any biasing of the signal power spectrum.

The aim of this work is to explore the effects of pixel masking on intensity mapping surveys in detail.  We perform our study using simulated intensity maps.  \citet{vtl} used N-body dark matter simulations for their analysis, but this technique is more numerically intensive than required for our purposes.  We therefore describe below a different method for simulating intensity maps, in which we assume a model for the matter power spectrum is known \emph{a priori}.  This makes it possible for us to quickly simulate relative large areas of the sky.  We use the matter power spectrum from CAMB \citep{camb} along with some empirically estimated luminosity functions to generate signal and foreground intensity  maps.

Using these simulations, we will demonstrate below that pixel masking is an effective technique for removing foregrounds when the signal is considerably brighter on average than the foregrounds, or when the pixel size of the survey is small enough that individual foreground sources can be isolated effectively.  We find that CO intensity maps meet the first criterion, and $\lya$ intensity maps presumably meet the second.  CII maps on the other hand, likely will not meet either criterion, and therefore cannot be easily cleaned using simple pixel masking.

It is important to note that \citet{BKK} (hereafter BKK) found that the amplitude of a CO intensity mapping power spectrum is extremely uncertain, and that different models yield very different results.  We have every reason to believe that the modeling of each and every signal and foreground line we discuss here is similarly uncertain.  Therefore it should be noted that the details of the luminosity function modeling described in this paper should be taken with a grain of salt and the exact amplitudes of the power spectra discussed below could vary significantly from the values we use. Nevertheless, our intention is to explore the effects of pixel masking on contaminated intensity maps, and our general conclusions should hold for many different models.

This paper is organized as follows.  Section 2 describes the method for calculating a power spectrum for an intensity map, then summarizes the models used for the three signal lines we consider.  Our simulation methods are explained in Section 3, first for maps with no large scale structure, and then for maps with clustering included.  Section 4 explains our method of removing foreground lines from these maps through pixel masking,  and Section 5 shows the results of this masking.  These results and pertaining issues are then discussed in Section 6, and we conclude in Section 7.

\section{Signal \& Foreground Models}
For all of the lines we discuss in this paper it is reasonable to assume that all of the emission in a given line comes from within individual galaxies.  We also assume that the galaxy luminosities in a given line are uncorrelated with one another.  Under these assumptions, the three dimensional power spectrum of an intensity map takes a fairly simple form
\be
P(k,z)=\tavg^2(z)P_{\rm{gal}}(k,z)+P_{\rm{shot}}(z).
\ee
The first term in this expression contains the sky-averaged brightness temperature $\tavg$ of the line, multiplied by the galaxy power spectrum $P_{\rm{gal}}$, which is the linear dark matter power spectrum multiplied by a bias factor (BKK).  This term, which we refer to as the clustering term, gives the contribution to the power spectrum from the large scale structure of matter in the universe.  If the intensity field we measure were spatially continuous this would be the entire power spectrum.  However, since the signal comes from a large number of randomly placed discrete sources, we must include an additional scale-independent shot noise term $P_{\rm{shot}}$.  

The values of $\tavg$ and $P_{\rm{shot}}$ depend on the astrophysical conditions within the emitting galaxies.  We quantify this astrophysical information through the luminosity function $\Phi(L)$, which gives the comoving number density of halos with luminosities in the desired line between $L$ and $L+dL$.  In terms of $\Phi(L)$, the average temperature and shot noise are
\be
\left<T\right>(z)=\frac{1}{8\pi}\frac{(1+z)^2}{\nu_{\rm{em}}^3}\frac{c^3}{k_bH(z)}\int_{L_{\rm{min}}}^{L_{\rm{max}}}L\Phi(L)dL,
\label{tavg}
\ee
and
\be
P_{\rm{shot}}(z)=\left[\frac{1}{8\pi}\frac{(1+z)^2}{\nu_{\rm{em}}^3}\frac{c^3}{k_bH(z)}\right]^2\int_{L_{\rm{min}}}^{L_{\rm{max}}}L^2\Phi(L)dL,
\label{shotnoise}
\ee
where $\nu_{\rm{em}}$ is the emission frequency of the line, $c$ is the speed of light, $k_b$ is Boltzmann's constant, and $H(z)$ is the Hubble parameter.  For some lines, the luminosity function can be difficult to measure, and it is often easier to estimate a relation $L(M)$ between the luminosity of a halo and its mass.  In this case, the integrals in equations (\ref{tavg}) and (\ref{shotnoise}) are over mass instead of luminosity, with the integrands replaced by $L(M)dn/dM$ and $L^2(M)dn/dM$.  In this case, we use the mass function $dn/dM$ from \citet{tinker} instead of the luminosity function to estimate the number of galaxies with a given luminosity.

Real intensity mapping surveys will survey the sky in several frequency bands, which corresponds to several different redshift slices.  The intensities in these slices will be correlated due to the existence of line-of-sight Fourier modes.  However, we can obtain a reasonable approximation of a true survey by treating each frequency band as an independent map, which can then be stacked with the others to improve signal to noise.  In order to facilitate the comparison of maps at different redshifts, we define our maps in angular coordinates and study the angular power spectrum $C_\ell$ in a single slice.  If we collapsed our whole 3D volume down to 2D, then we would lose a large amount of the information in our map.  Discussing the angular power spectrum of individual slices, however, only sacrifices the information present in the line-of-sight modes.  This is the approximation used in \citet{pullen} and \citet{BKK}.  All of the maps discussed in this paper should be thought of as a single slice of a full 3D survey.

Since intensity mapping experiments are likely to have fairly narrow frequency bands, it is reasonable to assume that the quantity $\tavg$ does not change significantly over the width of a single band.  This means that we can write the angular power spectrum as
\be
C_\ell(z)=\tavg^2(z)C_\ell^{\rm{gal}}+C_\ell^{\rm{shot}}.
\label{cl}
\ee
The galaxy angular power spectrum is calculated from $P_{\rm{gal}}$ using 
\be
C_\ell=\frac{2}{\pi}\int k^2P(k)\left[\int f(r)j_\ell(kr)dr\right]^2dk,
\ee
where $r$ is the comoving distance, $j_\ell(kr)$ is the spherical Bessel function and $f(r)$ is the selection function which is determined by the \citet{width of a single frequency band}.  For simplicity, we assume a top hat $f(r)$.  This integral is computationally difficult to evaluate, so we approximate it in the high-$\ell$ limit using the Limber approximation \citep{limber,rubin} and in the low-$\ell$ limit by assuming that $f(r)$ is a delta function.  More details on these approximations can be found in Section 2 of BKK.  The shot noise spectrum $C_\ell^{\rm{shot}}$ is scale independent, so the Limber approximation can be used on all scales.

The values of $\tavg$ and $P_{\rm{shot}}$ for a given line depend sensitively on the exact shape of the luminosity function of that line.  This luminosity function in turn depends on the conditions within the emitting galaxies which are highly uncertain, especially for high redshift sources.  We can attempt to predict $L(M)$ or $\Phi(L)$ for different lines using various empirical observations.  However, BKK found that different models assumed for CO emission yielded power spectra which spanned roughly two orders of magnitude in amplitude.  Therefore the models we summarize below should not be interpreted as precise predictions of the power spectra for these lines. Rather, they are intended as a means to gain understanding of how the {\it shapes} of the luminosity functions of foreground lines can affect how strongly they contaminate a target line, separately from the overall amplitude ambiguity.

\subsection{CO}

When modeling the emission of CO and its foreground lines, we assume that only halos with masses above some cutoff mass $M_{\rm{min}}$ (assumed here to be $10^9\ M_{\sun}$) can emit the line in question, and we assume that only a fraction $\fduty$ equal to the timescale of star formation over the age of the universe of halos are emitting the line at any given time.  For generality, we assume that the luminosity of a halo is a power law in its mass
\be
\frac{L}{L_{\sun}}=A\fduty\left(\frac{M}{M_{\sun}}\right)^b,
\label{Ab}
\ee
where $A$ and $b$ are free parameters.

In \citet{pullen} and \citet{lidz}, the calculation of $L(M)$ for CO uses a series of empirical scaling relationships, starting with a relation between the line luminosity and FIR luminosity of the form
\be
\frac{L_{\rm{FIR}}}{L_{\sun}}=C_{\rm{FIR}}\left(\frac{L'_{\rm{line}}}{\rm{K\ km\ s^{-1}\ pc^2}}\right)^{X_{\rm{FIR}}},
\ee
where $C_{\rm{FIR}}$ and $X_{\rm{FIR}}$ are constants set through observations.  In the above relation, the line luminosity is given in units commonly used for spectral line observations.  The conversion to solar luminosities is given by
\be
\frac{L_{\rm{line}}}{L_{\sun}}=3\times10^{-11}\left(\frac{\nu_{em}}{1\ \rm{GHz}}\right)^3\frac{L'_{\rm{line}}}{\rm{K\ km\ s^{-1}\ pc^2}},
\ee
\citep{car}.  These relations, in combination with the star formation rate-FIR luminosity and halo mass-star formation rate relations given in \citet{pullen}, we get the following expressions for $A$ and $b$:
\be
A=3\times10^{-11}\left(\frac{6.5\times10^{-8}}{C_{\rm{FIR}}}\right)^{1/X_{\rm{FIR}}}\left(\frac{\nu_{em}}{1\ \rm{GHz}}\right)^3,
\ee
\be
b=\frac{5}{3}X_{\rm{FIR}}^{-1}.
\ee
For CO, we follow BKK and \citet{pullen} in using the \citet{wang} CO-FIR relation with $C_{\rm{FIR}}=1.4\e{-5}$ and $X_{\rm{FIR}}=5/3$.  This gives the values $A=2\e{-6}$ and $b=1$ from \citet{pullen}.

When considering possible foreground lines, we consider only lines with lower rest frame frequencies than our target lines.  This is because, as shown in G14, projecting two power spectra from different redshifts boosts the amplitude of the lower redshift spectrum relative to the higher redshift one.  Because we are working here with angular power spectra instead of three dimensional power spectra, this projection effect is included naturally in our calculations.  For a CO survey targeted at $z=3$, we are therefore concerned with lines that have emission frequencies between the 115 GHz CO rest frame frequency and the 28.8 GHz observing frequency.  

The CO(1-0) line is expected to be considerably brighter on average than any other line in this range, i.e. $\tco$ should be much greater than $\tavg$ for any foreground line.  If we assumed that $L(M)$ were linear in $M$ for all lines this would mean that no foreground line could dominate over the CO signal.  However, the addition of the power law dependence on $M$ means that the CO map could still be contaminated by the brightest sources for a foreground line which has $b>1$.  The shot noise term in particular is very sensitive to the value of $b$.  For more general luminosity functions than the one we use here, this would mean that a line is potentially a problem if it falls off more slowly at high luminosities than the target line.  

We consider foreground lines emitted by four molecules: HCN, HCO$^+$, CN, and CS.  All of these molecules have higher critical densities than CO, and thus tend to trace denser regions of galaxies.  For each molecule, we use an empirical correlation with FIR to estimate $A$ and $b$.  The results and sources for each line are given in Table \ref{linetable}, along with those of CO for reference.  Note that for the  CN and CS lines the FIR relations used were measured for higher order transitions (3-2 and 7-6, respectively).  Thus we have made the assumption that the intensities of these lines are independent of which transition is being considered.  The situation in reality is likely not so simple, but this approximation will suffice for our purposes.  

\begin{table*}
\centering
\caption{Various parameters for the target CO(1-0) line as well as several possible foreground lines.  Parameters include the emission frequency $\nu_{em}$, the parameters of the FIR correlation $C_{\rm{FIR}}$ and $X_{\rm{FIR}}$, the $L(M)$ parameters $A$ and $b$, and the observable parameters $\tavg$ and $C_\ell^{\rm{shot}}$.}
\begin{tabular}{c|cccccccc}
\hhline{=========}
Line & $\nu_{em}$ (GHz) & $C_{\rm{FIR}}$ & $X_{\rm{FIR}}$ & $A$ & $b$ & $\tavg$ ($\mu$K) & $C_\ell^{\rm{shot}}$ ($\mu$K$^2$)  & Source \\ \hline
CO(1-0) & 115 & 1.35$\e{-5}$ & 1.67 & 2$\e{-6}$ & 1 & 0.60 & 7.8$\e{-7}$ & P13 \\
HCN(1-0) & 88 & 794 & 1.0 & 1.7$\e{-15}$ & 1.67 & 0.023 & 2.1$\e{-6}$ & \ \citet{HCN} \\ 
CN(1-0) & 113 & 1.6$\e{4}$ & 0.89 & 6.9$\e{-18}$ & 1.87 & 0.010 & 1.0$\e{-7}$ & \citet{CN} \\
CS(1-0) & 49 & 2.1$\e{4}$ & 1.0 & 2.7$\e{-18}$ & 1.67 & 2.1$\e{-4}$ & 3.8$\e{-8}$ & \citet{CS} \\
HCO$^+$ & 89 & 158 & 1.11& 1.2$\e{-15}$ & 1.5 & 1.94$\e{-4}$ & 3.4$\e{-11}$ & \citet{HCO} \\
\hhline{=========}
\end{tabular}
\label{linetable}
\end{table*}

\begin{figure}
\centering
\includegraphics[width=\columnwidth]{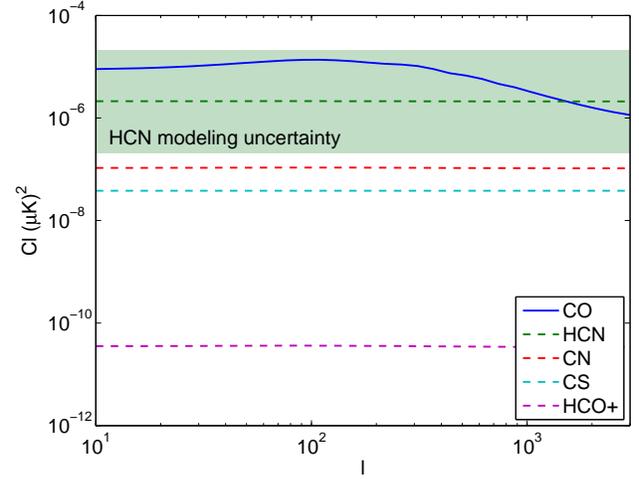}
\caption{Power spectra for lines described in Table \ref{linetable}.  The solid blue line is the target CO power spectrum, and the dashed lines show the spectra for the foreground lines.  Note that all of the foregrounds are shot noise dominated, and only HCN is comparable to CO.  The shaded region shows a rough estimate of the theoretical uncertainty in these power spectra.}
\label{foregrounds}
\end{figure}

Figure \ref{foregrounds} shows the clustering and shot noise power spectra for each of these lines compared with that of CO.  Because all of the foreground lines have $b>b_{\rm{CO}}$, they are all dominated by shot noise.  A possible qualitative reason for this is the fact that all of these lines tend to trace denser gas than CO, and thus their luminosities may be more sensitive to the environments in their host galaxies.

 Of the lines we consider, all but HCN have power spectra well below that of CO.  The HCN line however, actually starts to dominate over CO at $\ell\sim1000$.  Though HCN is only $\sim4\%$ as bright as CO on average, it has a higher value of $b$, which means that it produces a small number of bright sources which contribute a large amount of shot noise.  This demonstrates how a line with relatively small average intensity can still be a problematic foreground.  For the remainder of this paper, we will consider only the HCN foreground line since in our modeling the other lines are subdominant.

The shaded region in Figure \ref{foregrounds} shows a very crude estimate of the theoretical uncertainty in the HCN power spectrum, roughly an order of magnitude.  BKK found that the amplitude of the CO spectrum could also vary by roughly an order of magnitude in either direction.  For convenience we only show this shading for HCN, but all of the other spectra are similarly uncertain, if not more.  Given the vast amount of uncertainty both in the modeling of these spectra as well as in the empirical measurements used in our model, we do not attempt to make a more accurate estimate of the error bars on these spectra.

\subsection{Ly$\alpha$}
For the Lyman $\alpha$ line we consider a hypothetical survey targeted at $z=7$, and we follow the modeling of G14 and references therein, which we summarize briefly here.  Since the Ly$\alpha$ line will be observed in the infrared rather than the radio, we use intensity units here instead of brightness temperature.  Recombination and collision processes within galaxies give a $\lya$ luminosity
\be
\frac{L_{\lya}(M,z)}{L_{\sun}}=5.1\e8f_{\lya}(z)\left[1-f^{\rm{ion}}_{\rm{esc}}(M)\right]\frac{SFR(M,z)}{M_{\sun}\ \rm{yr}^{-1}},
\ee
where $f_{\lya}=3.34\e{-3}\times(1+z)^{2.57}$ is the fraction of $\lya$ photons not absorbed by dust, and $f^{\rm{ion}}_{\rm{esc}}=\exp\left(-5.18\e{-3}\times M^{0.244}\right)$ is the escape fraction of ionizing photons.  The star formation rate is parameterized by
\begin{multline}
SFR(M,z=7)=1.6\e{-26}\left(\frac{M}{M_{\sun}}\right)^{2.59}\left(1+\frac{M}{M_1}\right)^{-0.62} \\ \times\left(1+\frac{M}{M_2}\right)^{0.4}\left(1+\frac{M}{M_3}\right)^{-2.25}.
\end{multline}
We take $M_{\rm{min}}=10^{8}\ M_{\sun}$ and $M_{\rm{max}}=10^{13}\ M_{\sun}$.
Note that for simplicity we neglect $\lya$ emission from the IGM.  The IGM contribution is small compared to the halo emission, and it is not as easily simulated using our methods.  We calculate the bias for Ly$\alpha$ following G14.

\begin{figure}
\centering
\includegraphics[width=\columnwidth]{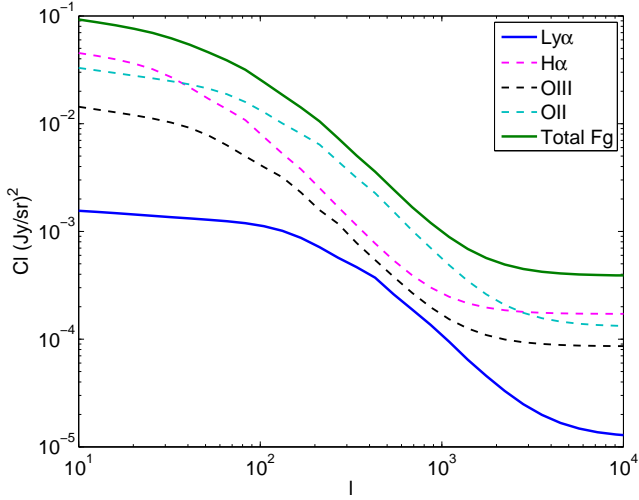}
\caption{Theoretical power spectra for $\lya$ and its three foregrounds as well as the total foreground spectrum}
\label{lyafgps}
\end{figure}

For $\lya$ at $z=7$, three foreground lines are considered: H$\alpha$ coming from $z\sim0.5$, OIII from $z\sim0.9$, and OII from $z\sim1.6$.  For these lines, the luminosity function is assumed to be a Schechter function
\be
\Phi(L)dL=\phi_*\left(\frac{L}{L_*}\right)^\alpha\exp\left(-\frac{L}{L_*}\right)\frac{dL}{L_*},
\ee
where $\Phi(L)$ is the comoving number density of halos with luminosities between $L$ and $L+dL$, and $\phi_*$, $L_*$, and $\alpha$ are parameters which we obtain from \citet{Ly}.  With the above relations, we can calculate angular power spectra for $\lya$ and the three foreground lines using equation (\ref{cl}).  Following G14, we assume that the halo bias for these lines is proportional to the halo mass.  Figure \ref{lyafgps} shows the calculated power spectra for $\lya$ and the three foreground lines along with the total foreground spectrum.  Note that the foreground lines dominate entirely over the signal, and unlike CO, the foregrounds have a significant clustering component.

\subsection{CII}
We base our simulations of CII intensity maps on the modeling done by \citet{S14}, hereafter S14.  We consider a possible survey targeted at $z=7$.   For the CII luminosity we use model $\mathbf{m_2}$ from Table 1 of S14, where the luminosity is given by
\be
\log10\left(\frac{L_{\rm{CII}}}{L_{\sun}}\right)=\log10\left(\frac{\psi(M,z)}{M_{\sun}}\right)+6.9647,
\ee
where
\be
\psi(M,z)=M_0\left(\frac{M}{M_a}\right)^{a_{\rm{CII}}}\left(1+\frac{M}{M_b}\right)^{b_{\rm{CII}}}.
\ee
The parameters in the formula for $\psi$ at $z=7$ are $M_0=6.6\e{-6}\ M_{\sun}$, $M_a=10^8\ M_{\sun}$, $M_b=1.6\e{11}\ M_{\sun}$, $a_{\rm{CII}}=2.25$, and $b_{\rm{CII}}=-2.3$.  The average CII intensity and shot noise can then be calculated using equations (\ref{tavg}) and (\ref{shotnoise}) above.  

\begin{figure}
\centering
\includegraphics[width=\columnwidth]{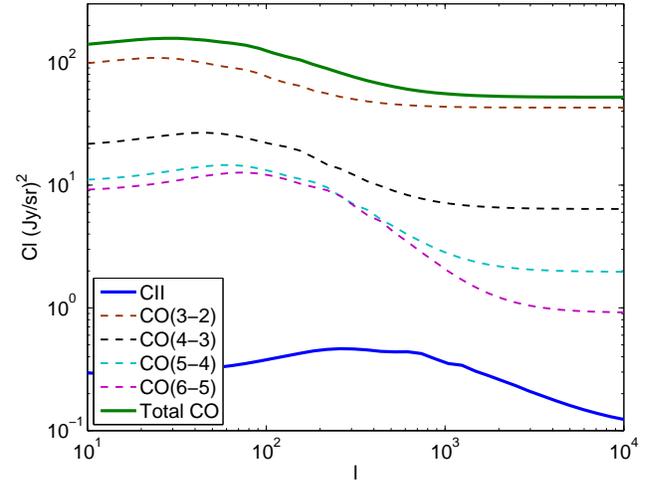}
\caption{Theoretical power spectra for CII and its four foreground CO lines along with the total foreground spectrum.}
\label{C2spec}
\end{figure}

The primary foregrounds for a CII survey come from CO.  Since such a survey would observe at roughly 240 GHz, we need not worry about the CO(1-0) line at 115 GHz or the CO(2-0) line at 230 GHz.  The higher order transitions are potentially problematic though.  We estimate the intensities of the higher order CO lines using our formulae above along with the line ratios given for submillimeter galaxies in Table 2 of \citet{cw}.  This table gives the ratio $L'/L'_{CO(1-0)}$ for the transitions from CO(2-1) to CO(5-4).  Following S14, we consider the CO(6-5) line as well, and assume that it has the same luminosity ratio as the CO(5-4) transition.  We also calculate the bias by assuming that the halo bias is proportional to the halo mass.  Figure \ref{C2spec} shows the calculated power spectra for these CO lines as well as the spectrum of the target CII line.   As with $\lya$, the foreground lines are strongly clustered and considerably brighter than the target line.

\section{Simulations}
In order to explore the effects of foreground contamination in more detail it is useful to create simulated maps of various lines.  In this section we describe our simulation methods, first for maps with shot noise power spectra, then for maps with large scale clustering included.  We will explain our simulation process using CO at $z=3$ as an example.

\subsection{Shot noise maps}
When simulating a map, we first define a grid of pixels, with the solid angle of each pixel defined by the angular resolution of a hypothetical instrument, the pixel depth by the spectral resolution, and the total area defined by the proposed survey area.  For our CO simulations, we use a fiducial survey observing at 30 GHz (corresponding to CO at $z=3$) covering 550 sq. deg. with an angular resolution of 10 arcmin and $\delta\nu/\nu=10^{-3}$.  These numbers are similar to those discussed in \citet{pullen} and BKK.  Since each pixel in this simulated map is large compared to a galaxy, we can assume that the line emitters are essentially point sources.  The expected mean number of sources in each pixel is given by
\be
\Nm(z)=nV_{\rm{pix}}\fduty=V_{\rm{pix}}(z)\fduty(z)\int_{\Mmin}^\infty\frac{dn(z)}{dM}dM,
\ee
where $n$ is the total halo number density and $V_{\rm{pix}}$ is the comoving volume of a single pixel.

With no clustering taken into account, galaxies should be Poisson distributed on the sky.  Thus we can draw the number of sources in each pixel from a Poisson distribution with mean $\Nm$.  Assuming the line luminosities of individual sources are uncorrelated, we can then randomly assign a luminosity to each source.  Since we are using a simple model in which luminosity is determined by mass, we first draw a mass from the distribution 
\be
P(M)dM = \frac{1}{n}\frac{dn}{dM},
\label{PM}
\ee
then calculate the line luminosity from $L(M)$ for each galaxy and convert this to a brightness temperature to get the finished map.  

Since we are simulating relatively small regions of the sky, we can calculate power spectra in the flat sky approximation, similar to the method used in \citet{cc}.  For simplicity, we assume the survey area is a square located near the equator of whatever sky coordinates are in use.  

Consider a map with $N_{\rm{pix}}$ pixels at positions $\vx=(x_i,y_i)$, where $x_i$ and $y_i$ run from 1 to $\sqrt{N_{\rm{pix}}}$.  The intensity at each pixel $T(\vx)$ can be decomposed into Fourier modes $a_\vk$ through the discrete Fourier transform,
\be
a_\vk=\frac{1}{N_{\rm{pix}}}\sum_\vx T(\vx)e^{2\pi i\vx\cdot\vk/N_{\rm{pix}}}.
\ee
The angular power spectrum of the map is then
\be
C_{\ell=2\pi k/\sqrt{\Omega}}=\Omega C_k,
\ee
where $\Omega$ is the total solid angle of the survey and
\be
C_k=\left<\left|a_\vk\right|^2\right>.
\label{ck}
\ee
The average in equation (\ref{ck}) is taken over all k modes in the interval $k-1/2\leq|\vk|<k+1/2$.  Note that in this calculation $\vx$ and $\vk$ are in units of pixels and pixels$^{-1}$ respectively.

\begin{figure*}
\centering
\includegraphics[width=\textwidth]{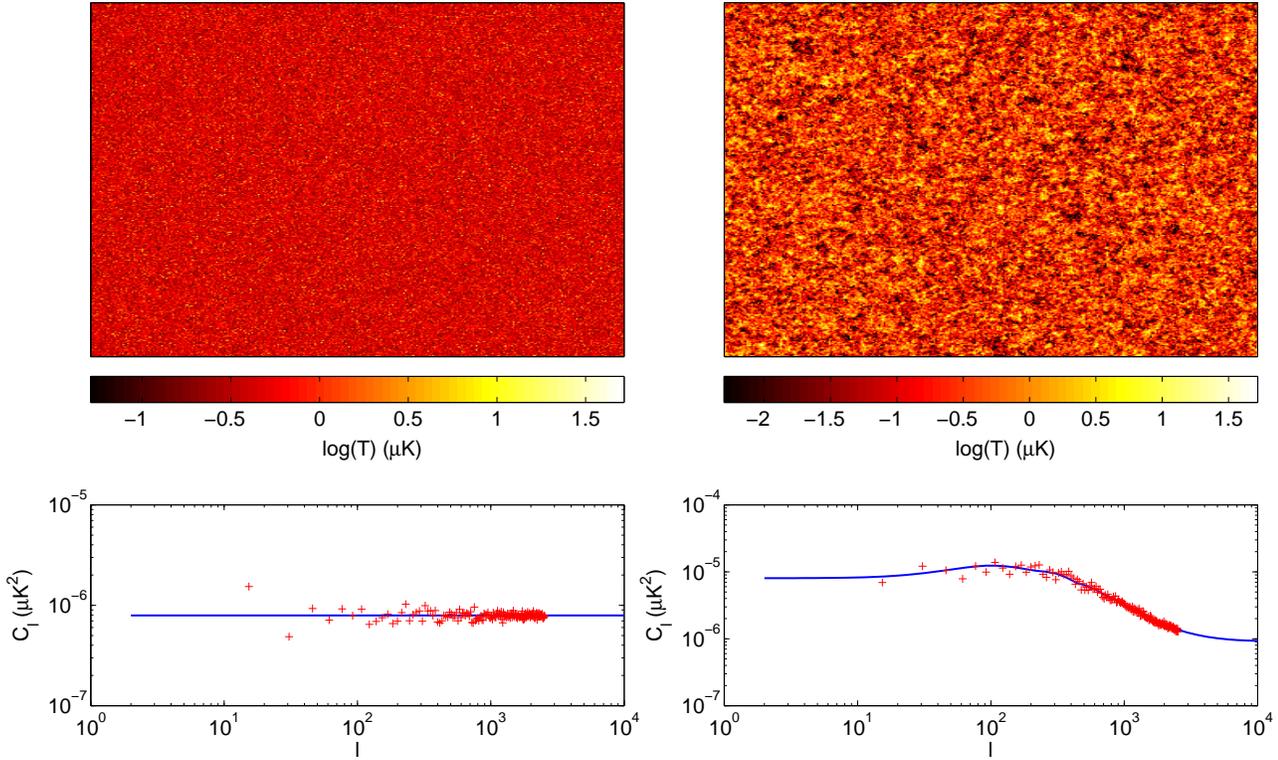}
\caption{(Top left) Simulated shot noise dominated CO intensity map from $z=3$.  (Bottom left) Power spectrum of simulated map (red points) overplotted with the predicted shot noise power spectrum from equation (\ref{shotnoise}).  (Top right) Simulated CO intensity map from $z=3$ with clustering included. (Bottom right) Power spectrum of simulated clustering map overplotted with the predicted power spectrum including both clustering and shot noise components from equation (\ref{cl}).}
\label{COsim}
\end{figure*}

The left hand column of Figure \ref{COsim} shows a map of CO emission simulated using the above technique as well as its power spectrum.  The power spectrum has the expected scale-independent form of a shot-noise dominated sample.  The solid line is the predicted shot noise amplitude calculated using equation (\ref{shotnoise}) for CO at $z=3$, included to show the consistency between our model and our simulations.  The scatter of the points around this line is cosmic variance error due to the finite size of our simulation.  There is no instrumental noise included in the simulated map.

\subsection{Adding clustering}
The simulations above assumed that galaxies are randomly distributed on the sky, but in reality the large scale structure of the universe imposes a pattern of clustering on the galaxy distribution.  The typical method of simulating maps with realistic clustering is to perform an N-body dark matter simulation and populate the result with galaxies \citep{vtl}.  However, this is computationally intensive and makes it difficult to simulate large sky areas.  Since we have predictions for the angular power spectra of these maps we choose to generate random density fields with the desired power spectra rather than attempt to simulate the entire history of the region being simulated.  

We want to simulate a density field $\delta(\vx)=(n(\vx)-\bar{n})/\bar{n}$ which includes clustering.  The $\delta$ map should have zero mean and its power spectrum should be the halo power spectrum.  In a  pixel located at $\vx$, instead of drawing the number of galaxies from a Poisson distribution with mean $\Nm$, we use $\Nm(\delta(\vx)+1)$.  Since galaxy luminosities do not depend on the clustering, the procedure for assigning luminosities is the same as before.  This will give a map with a power spectrum that contains both clustering and shot noise components.

It only remains then to generate a density field with the desired halo power spectrum.  If the field is Gaussian, the process is fairly straightforward.  In a full-sky Gaussian random field, each spherical harmonic coefficient $a_{\ell m}$ is drawn from a Gaussian distribution with variance $\sqrt{C_\ell}$.  In the flat sky approximation, we draw each Fourier mode with magnitude in the range $k-1/2\leq|\vk|<k+1/2$ randomly from a Gaussian with variance $C_k$.  This map can then be converted back to spatial coordinates using the inverse discrete Fourier transform.  One important thing to note is that the resulting map must be real, while the Fourier modes are complex.  This means that we must impose the condition that $a_{-\vk}=a_\vk^*$.  

Unfortunately, at the redshifts we are considering, the galaxy distribution is highly non-Gaussian.  Attempting to impose a Gaussian distributed $\delta(\vx)$ produces pixels where $\Nm(\delta(\vx)+1)$ is negative, which is obviously unphysical.  A better approximation for the galaxy distribution would be to use a log-normal distribution \citep{cj}.  This distribution has the important property that it is zero for negative densities and it appears to be a reasonably good fit to the observed galaxy distribution.  The procedure for generating a log-normal random field uses the fact that that a log-normal map $\delta_{LN}(\vx)$ can be generated from a Gaussian map with variance $\sigma_G^2$ using
\be
\delta_{LN}(\vx)=e^{\delta_G(\vx)-\sigma_G^2/2}-1.
\label{deltaconv}
\ee
There exists a convenient relation between the correlation functions $\xi(r)$ of two maps related in this way
\be
\xi_G(r)=\ln\left[1+\xi_{LN}(r)\right].
\label{GLN}
\ee
This allows us to generate a log-normal random field $\delta_{LN}$ with the desired characteristics.  

We start with the galaxy power spectrum $C_\ell^{\rm{gal}}$ and convert it to a flat sky approximation $C_k^{\rm{gal}}$.  From this, we can calculate the correlation function we want our log-normal field to have using
\be
\xi_{LN}(r)=\frac{1}{2\pi^2}\int_0^\infty k^2C_k^{\rm{gal}}\frac{\sin(kr)}{kr}dk.
\ee
We then calculate $\xi_G$ using equation (\ref{GLN}) and convert this to a new angular power spectrum using
\be
C_k^G=\frac{2}{k}\int_0^\infty r\xi_G(r)\sin(kr)dr.
\ee
We can then draw a Gaussian random field $\delta_G$ with this power spectrum using the procedure outlined above, then convert it to a log-normal random field using equation (\ref{deltaconv}).

The result of this process is a map like the one shown in the top right panel of Figure \ref{COsim}, with a power spectrum as shown in the bottom right panel.  It is obvious when comparing these plots to those in the left hand column that the second map is much more strongly clustered and that the power spectrum contains a distinct scale-dependent component.  The blue curve in this plot shows the predicted CO power spectrum from Section 2.1.  This predicted spectrum is in good agreement with the simulated data, which again shows the consistency between our theoretical models and our simulations.

\subsection{Simulation results}

\begin{figure}
\centering
\includegraphics[width=\columnwidth]{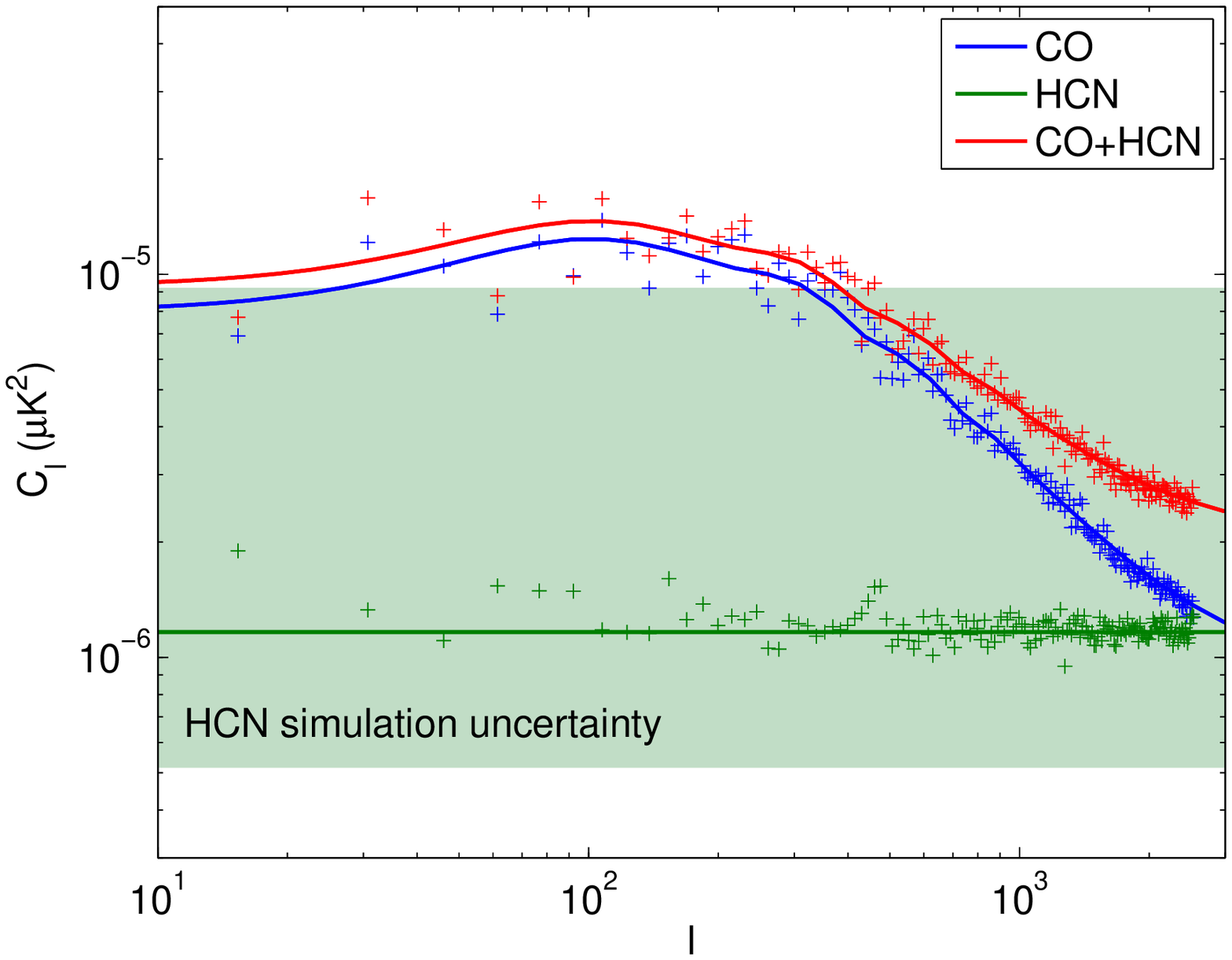}
\caption{Power spectra of simulated maps of CO at $z=3$ (blue), HCN at $z=2$ (green), and the sum of the two (red).  Solid curves are best fit power spectra allowing $\left<T\right>$ and $C_\ell^S$ to vary.  As expected, the HCN foreground contributes a significant amount of extra shot noise at high $\ell$. The shaded region corresponds to the 95\% confidence interval for the simulated HCN maps.  Variations from simulation to simulation can cause the power spectrum amplitude to change significantly.}
\label{fgps}
\end{figure}

With this method in place, we can now simulate maps for any line for which we can define either a function $L(M)$ or a luminosity function $\Phi(L)$.  For lines where we use a luminosity function, we simply replace the mass function in equation (\ref{PM}) with $\Phi(L)$. Since we are working in angular space, it is also very easy to combine maps for different lines.  A map containing two lines is simply the sum of the maps for the individual lines.  If we simulate a map for CO at $z=3$ and a foreground HCN line at $z=2$, we get power spectra like the ones shown in Figure \ref{fgps}.  The solid curves in this plot are the best fit power spectra obtained by treating $\left<T\right>$ and $C_\ell^S$ as free parameters.  As expected, the effect of the foreground contamination is to significantly increase the shot noise of the map, which is particularly obvious at high $\ell$.

The amplitudes of the power spectra in these simulations depend on exactly how many halos are drawn and exactly what masses are assigned to them.  This means that the result can vary somewhat from the theoretical predictions.  The HCN power spectrum is particularly sensitive to this, since it depends so much on the highest mass halos.  The shaded region in Figure \ref{fgps} shows the 95\% confidence range of the HCN power spectrum, calculated by comparing the results of 500 simulated HCN maps.

\begin{figure*}
\centering
\includegraphics[width=\textwidth]{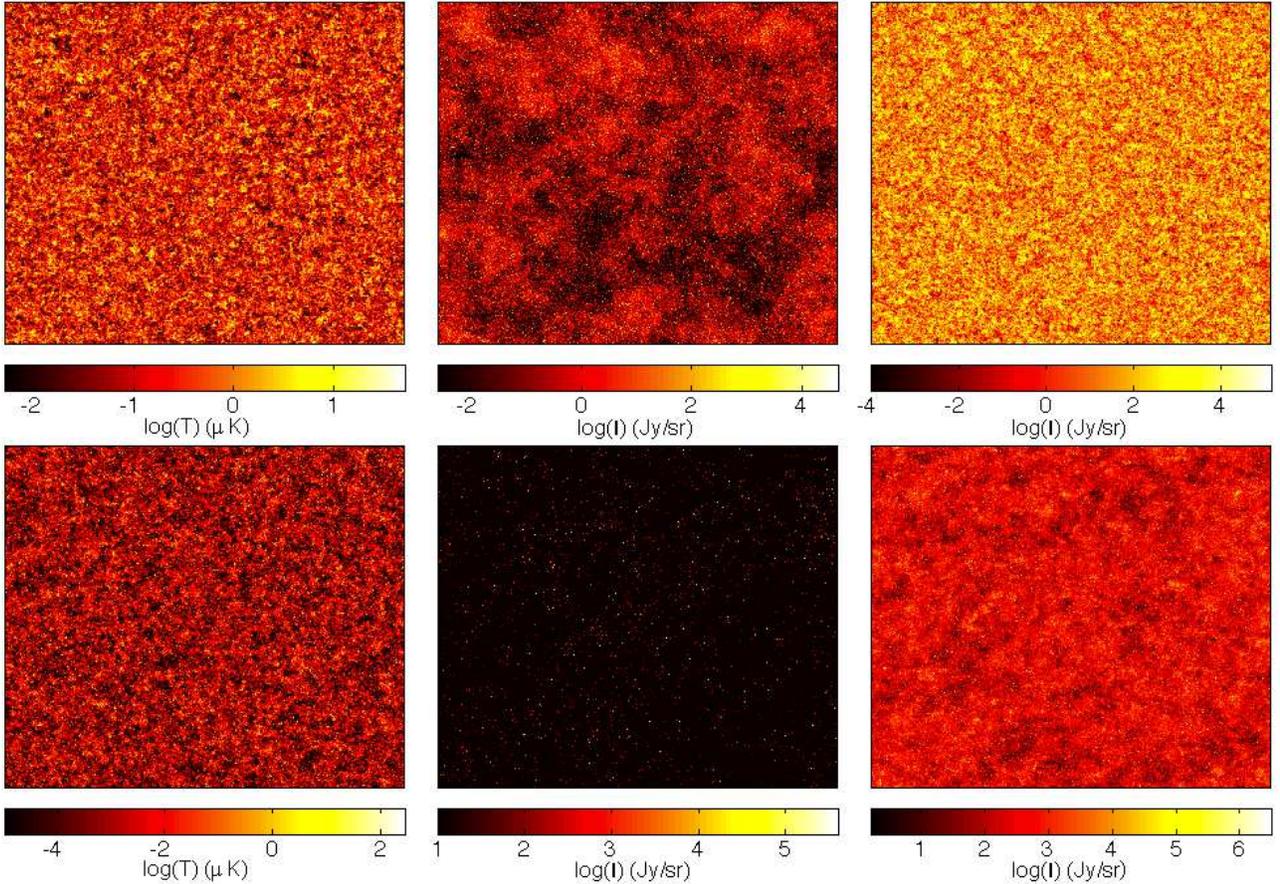}
\caption{Simulated maps of CO at $z=3$, $\lya$ at $z=7$, and CII at $z=7$ (top row, left to right) along with their foregrounds (bottom row).  The CO simulations cover 550 deg$^2$ with 10 arcmin resolution, the $\lya$ simulations cover 1 deg$^2$ with 0.1 arcmin resolution, and the CII simulations cover 100 deg$^2$ with 3.2 arcmin resolution.}
\label{allsims}
\end{figure*}

Figure \ref{allsims} shows simulated maps for CO at $z=3$ and $\lya$ and CII at $z=7$ along with their foregrounds in order to illustrate the qualitative differences between them.  The CO simulations cover 550 deg$^2$ with 10 arcmin resolution and $\Delta\nu/\nu_{\rm{obs}}=10^{-3}$, the $\lya$ simulations cover 1 deg$^2$ with 0.1 arcmin resolution and $\Delta\nu/\nu_{\rm{obs}}=1/40$ (G14), and the CII simulations cover 100 deg$^2$ with 3.2 arcmin resolution and $\Delta\nu/\nu_{\rm{obs}}=1.7\e{-3}$ (S14).

The HCN foreground map is considerably fainter on average than the CO map, but though the brightest sources are difficult to see on this image the colorbar shows that the HCN map extends to higher intensities than the CO map.  The $\lya$ maps cover a much smaller region of the sky than the others, so the clustering features in this image appear much larger.  In addition, each pixel in the $\lya$ foreground map covers a rather small volume of space, so most of the pixels in this map are dark.  Though the $\lya$ foregrounds are brighter than $\lya$ on average, most of the intensity in this map comes from a small number of pixels.  The foreground map for CII is much brighter than the signal, and the pixels in the CII map are large enough that there is strong emission in most pixels.

Though the models we use in this work are fairly simple, this simulation method is straightforward to generalize to different, more complex models.  Any power spectrum and luminosity function can be input to get a simulated map.  One important effect that we have not taken into account here is the correlation between maps made at different frequencies due to line-of-sight Fourier modes.  Since the target lines we consider here are widely separated from the foreground lines in frequency space this effect should not be significant for the problems discussed here.  However, it will need to be taken into account when attempting to accurately simulate the three-dimensional data taken by realistic intensity mapping experiments.

\section{Method}
\begin{figure}
\centering
\includegraphics[width=\columnwidth]{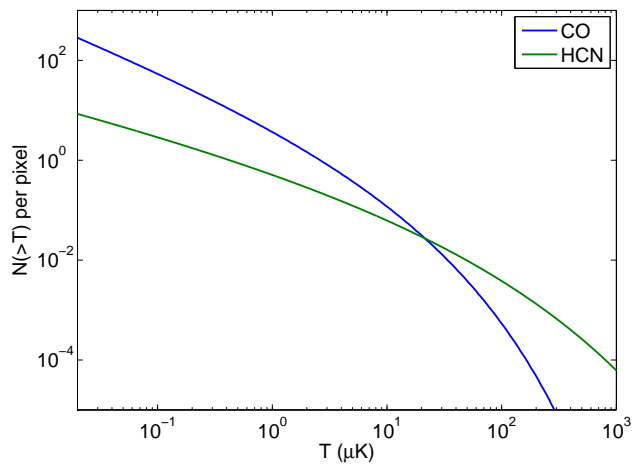}
\caption{Number of sources per pixel which contribute brightness temperature greater than $T$.  At low intensities, the target CO line dominates as expected.  However, at high intensities the HCN emitters begin to dominate because the HCN luminosity function does not fall off as quickly as that of CO.}
\label{Nabove}
\end{figure}

As noted above, the majority of the contamination in CO surveys comes from a few bright foreground emitters which add a large amount of shot noise to a survey.  As noted by G14, this means that the foreground effect could be mitigated if we simply mask out the brightest pixels in our survey.   Figure \ref{Nabove} shows the number of sources/pixel which produce brightness temperatures above a given value.  At brightness temperatures above a few hundred K, there are more HCN emitters than CO emitters.  This is primarily due to the fact that $b_{\rm{HCN}}>b_{\rm{CO}}$, with some additional contribution from the fact that there are more high mass halos at $z\sim2$ than at $z\sim3$.

This means that if a given pixel has a very high intensity, it is likely that the extra flux is coming from a foreground HCN emitter.  Thus if we mask all pixels brighter than a given value we will remove on average much more foreground emitters than target emitters.  In addition, the foreground sources we mask are exactly the bright sources which produce the worst contamination.  Therefore we expect that we can clean the foregrounds out of a map by masking out the brightest pixels.

The effects of pixel masking on the power spectra of these maps are difficult to understand analytically.  We can make a very rough estimate if we assume that we can perfectly isolate and remove all sources which contribute an intensity greater than some cutoff.  The ratio of the average brightness temperature of a line before and after applying this cutoff is
\be
\frac{\tavg_{\rm{masked}}}{\tavg_{\rm{unmasked}}}=\frac{\int_{M_{\rm{min}}}^{M_{\rm{cut}}}L(M)\frac{dn}{dM}dM}{\int_{M_{\rm{min}}}^{M_{\rm{max}}}L(M)\frac{dn}{dM}dM},
\label{analytic}
\ee
If we choose to remove $\sim1\%$ of the pixels from a map, $M_{\rm{cut}}$ is the mass of a halo such that $1\%$ of halos have $M>M_{\rm{cut}}$.  For CO, we have $M_{\rm{cut}}=6.2\e{10}\ M_{\sun}$ and for HCN we have $9.5\e{10}\ M_{\sun}$.  The ratio of the shot noise before and after masking is similar to equation (\ref{analytic}) with $L(M)$ replaced with $L^2(M)$.

This analytical procedure predicts that the ratio of the masked to unmasked CO intensity will be $\sim0.7$ and that of the masked to unmasked CO shot noise will be $\sim0.06$.  The ratio of masked to unmasked HCN shot noise is predicted to be $\sim2\e{-5}$.  Therefore, although masking pixels causes the CO power spectrum to decrease, we expect the HCN spectrum to decrease far more, leaving behind a signal dominated power spectrum.  However, the procedure described here neglects a few key effects which would be present in a real map.  If a map pixel contains a bright source, masking it will also remove all of the fainter sources present in the same pixel.  In addition, some pixels will be masked because they contain a large number of faint sources rather than a single bright one.  In order to fully account for these effects, we need to rely on our simulated maps, as we do below.

One question that arises naturally is exactly how many pixels should be masked?  For CO, and any other hypothetical line where all of the foregrounds are shot-noise dominated, the answer is simple.  We only need to mask until the masked power spectrum shows clustering behavior out to the desired angular scale.  Even if the CO foregrounds are faint enough that CO dominates entirely, this masking will reveal the CO clustering behavior on scales which are normally obscured by shot noise.  For lines with clustered foregrounds, the answer is less clear.  If we have a reasonable estimate of the luminosity functions of the signal and foregrounds, we can use a plot like Figure \ref{Nabove} to predict a cutoff intensity (this is the method used in G14). 

\section{Results}
Here we show the results of applying the masking procedure described above to CO, $\lya$, and CII intensity maps.

\subsection{CO}

\begin{figure}
\centering
\includegraphics[width=\columnwidth]{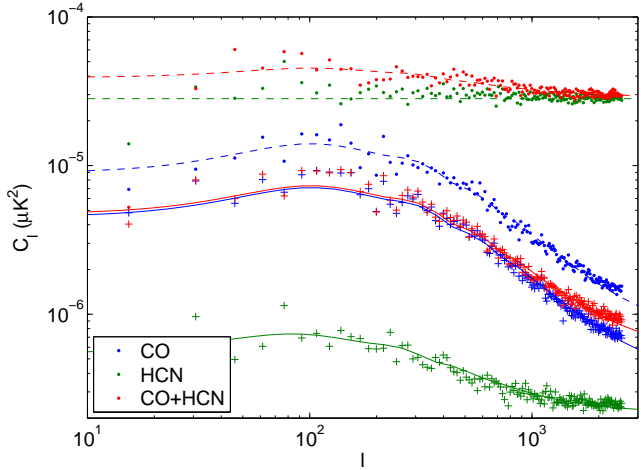}
\caption{Power spectra and best fit curves for simulated maps of CO at redshift 3 (blue), HCN at redshift 2 (green) with the amplitude boosted to dominate over CO, and the sum of the two (red).  Dashed curves/dots show the spectra before masking, solid curves/pluses show the results of masking all pixels above 7 $\mu$K ($\sim1\%$ of the pixels in the total map).  After masking, the maps with and without foregrounds have very similar power spectra.}
\label{mask}
\end{figure}

Figure \ref{mask} shows the effect of pixel masking on the power spectra of three simulated maps: one with just CO, one with HCN, and the sum of the two.  In order to make the effects of masking more obvious, we have simulated the HCN map with the value of the $A$ parameter increased by a factor of 3.  This creates a map where the foreground shot noise entirely dominates over the signal.  Given the uncertainties in the power spectrum modeling, it is not impossible for this to be the case in reality.

In Figure \ref{mask}, the dots and dashed curves show the power spectra and best fit curves for the three maps with no masking.  The boosted foreground power spectrum entirely dominates on all scales.  The pluses and solid curves show the spectra of the maps after all pixels brighter than 7 $\mu$K are masked.  This value corresponds to masking roughly 1\% of the pixels in the map with both signal and foregrounds.  After masking, the foreground power spectrum has dropped dramatically, and  the CO power spectrum has fallen by a much smaller amount.  But most importantly, the red total power spectrum is very similar to the CO power spectrum.  The power spectrum of the map with foregrounds is nearly identical to that of a map without them.  There remains some small amount of shot noise contamination, but this could be removed by choosing a lower cutoff value for masking.  Thus it appears that foreground contamination in CO intensity maps can indeed be mitigated by masking bright pixels.

It is worth mentioning that the amount by which the CO power spectrum drops differs somewhat from the simple prediction made in Section 4.  The clustering term in our simulation decreases roughly 30\% less than the simple calculation predicts, and the drop in shot noise is ten times less than predicted.  As mentioned above, this happens because the calculation in Section 4 does not take into account the fact that there are multiple sources in each pixel, which will either be masked along with a single bright source or add together to mimic a bright source.  This discrepancy makes it difficult to use equation (4.1) to predict the unmasked spectrum from a masked one.

\subsection{$\lya$}

\begin{figure}
\centering
\includegraphics[width=\columnwidth]{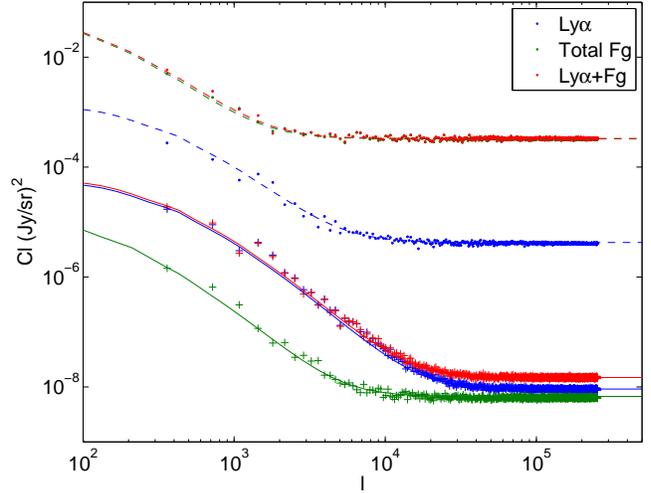}
\caption{Power spectra and best fit curves for simulated maps of $\lya$ at $z=7$ (blue), the sum of the three foreground lines (green), and all four lines combined (red).  Dashed curves/dots show the spectra before masking, solid curves/pluses show the results of masking all pixels above $\sim100$ Jy/sr (3\% of the pixels in the total map).  As with CO, the masking removes a large amount of the foreground contamination.}
\label{Lyamask}
\end{figure}

To facilitate comparison between our results and those given in G14, we choose to mask 3\% of the pixels in our $\lya$ simulations.  Figure \ref{Lyamask} shows the effectiveness of this masking.  The dashed lines show the power spectra of the signal, foregrounds, and total map before masking, and the solid lines show the spectra after masking.  For the sake of visibility, we have plotted only the total foreground spectra rather than those of the three individual lines.  After masking, the foreground has dropped below the signal and the map which includes both signal and foregrounds gives a spectrum very similar to the map without foregrounds.  Thus it appears that for the model and telescope resolution simulated here it is possible to remove most of the foreground contribution to the power spectrum.

\subsection{CII}

\begin{figure}
\centering
\includegraphics[width=\columnwidth]{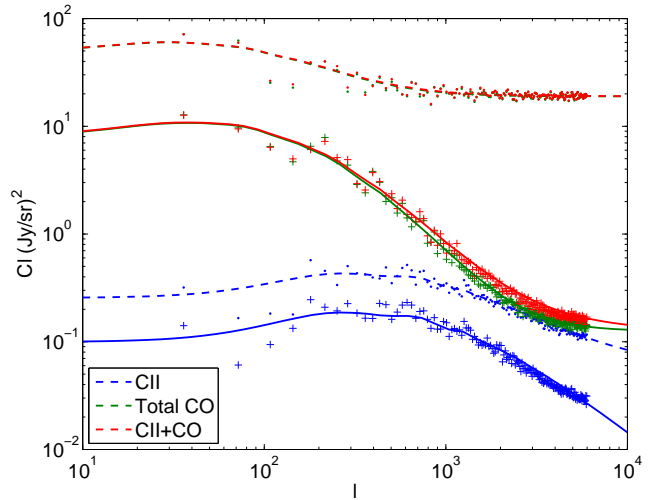}
\caption{Power spectra and best fit curves for simulated maps of CII at $z=7$ (blue), the total CO foreground (green), and the total emission (red).  Dashed curves/dots show the spectra before masking, solid curves/pluses show the results after masking all pixels brighter than $\sim10^4$ Jy/sr (1\% of the pixels in the total map).  After masking, the foregrounds still dominate over the signal.  The foregrounds continue to dominate no matter what masking percentage is chosen.}
\label{CIIcut}
\end{figure}

Figure \ref{CIIcut} shows the power spectra of the CII simulations before and after masking 1\% of the pixels.  Note that unlike in our CO and $\lya$ simulations, the foreground lines dominate over the signal even after masking.  Though we have plotted the results of masking $1\%$ of the pixels, the same basic result holds true no matter how many pixels we mask.  At this resolution, it is not possible to move the foreground power spectrum below that of the signal no matter how many pixels are removed.

\section{Discussion}

The details of masking the simulated maps explain why the three lines we consider behave somewhat differently under masking.  For the CO map, the contamination is shot noise from the few brightest galaxies, which is easily removed.  However, both $\lya$ and CII have foregrounds which include strong clustering components.  In addition, for both of these lines the average intensity of the foregrounds is large compared to that of the signal, as seen in G14 and S14.  Despite this, the masking works well for $\lya$ and poorly for CII.  The reason for this has to do with the size of the pixels used in each map.  $\lya$ is a higher frequency line than CII or CO, so it is easier to map with high angular resolution.  This means that it is easier to isolate a single bright foreground source without removing as much signal.  Our results show that if a CII survey is limited to arcminute scale pixels then foregrounds cannot be easily cleaned through pixel masking, and we must resort to cross-correlations to isolate the target line.

The drop in the amplitudes of the target spectra after masking is an unfortunate side effect of pixel masking.  Ideally, an intensity mapping survey would be able to recover the dashed blue curve in Figure \ref{mask} rather than the solid blue curve.  This is because we cannot avoid masking some CO sources when removing HCN.  As noted in \citet{vtl}, some of the information in the power spectrum is lost in this masking process, specifically that encapsulated in the amplitudes of the clustering and shot noise terms.  On a brighter note, since we are only masking a percent or so of the pixels, the shape of the clustering term of the power spectrum does not change significantly when masked.  This means that masking allows us to recover the \emph{shape} of the galaxy power spectrum even on scales where the foregrounds dominate over the signal.  

This is a potentially useful measurement for a variety of \emph{cosmological} purposes.  For example, \citet{21mg} show that modified gravity models alter the shape of the power spectrum which would be measured by a 21 cm survey.  Similar changes could be studied in the masked maps we describe here.  However, all of the \emph{astrophysical} information which was contained in the \emph{amplitudes} of the clustering and shot noise components is lost in masking, so we cannot make statements about the luminosity functions of the target galaxies.  This means that, while these intensity mapping surveys could still be useful with only this simple foreground cleaning, it may be necessary to use other foreground cleaning methods to reach their full potential.

Nonetheless, the information lost in pixel masking could potentially be recovered in a number of ways.   Unfortunately, the estimate of the change in power given in equation (4.1) differs significantly from our simulations, so this cannot be used to recover the unmasked power.  However, cross correlations between different maps are always possible, and S14 and \citet{vtl} discuss the possibility of using information in another wavelength to isolate pixels which contain foreground galaxies, thus limiting the amount of signal lost in masking.  G14 also state that the behavior of line-of-sight Fourier modes may be different for signal and foreground lines.  Another way could be to fit the one-point statistics of the map using P(D) analysis \citep{pofd} as the bright pixels are progressively removed, and compare the results with theoretical simulations to retrieve the original amplitudes.  This could be a means of obtaining the astrophysical information lost in masking (Breysse, Kovetz, \& Kamionkowski 2015).  Both the power anisotropy mentioned in G14 and the one point statistics could also be used to determine if the signal or a low-redshift foreground line dominates after a map has been masked.

Note that care must be taken when fitting uncleaned intensity mapping power spectra to determine cosmological or astrophysical parameters.  For example, The power spectrum of a CO intensity map can in principle be fit to determine the values of various model parameters (Li et al. 2015), and the fitted parameters will depend on how much shot noise is present in the map.  However, the foreground line will add an uncertain amount of extra shot noise to the spectrum, which will bias the results of any fit.

We have not included the effects of instrumental noise in any of these calculations.  Though we leave for future work a full analysis of the behavior of noisy maps with masked pixels, we can make some basic arguments to predict whether or not our basic premise would hold in a map with noise.  If a map has too much noise, then the pixels which are masked will be bright due to random noise fluctuations in no correlation with the brightness of the target or foreground galaxies in those pixels.  Therefore, if a simulation has too much noise this pixel masking technique is useless.  The survey parameters used in BKK, which were chosen to provide a reasonable chance of detecting CO at $z=3$, give a noise per pixel $\sigma_N=1.7\ \mu$K.  If we generate noise maps with the value in each pixel drawn from a Gaussian with zero mean and standard deviation of $\sigma_N$, we can estimate how many of the masked pixels will be bright due to noise rather than signal.  We find that adding noise to our simulated CO+HCN map increases the number of pixels above our $7\ \mu$K cutoff by roughly 25\%.  These spurious bright pixels will reduce the effectiveness of the masking, but it should still be possible to remove most of the bright sources from a map.  If the noise is significantly stronger, the masking will be ineffective.  However, in this case the signal to noise ratio for detecting the CO line at all becomes considerably smaller as well.

In addition, we assumed the linear form for the underlying matter power spectrum in our calculations.  Though we account for some nonlinearity by using a lognormal galaxy density field, a full treatment of the nonlinearities would add more power to our maps on small scales.  However, this would likely not have a significant qualitative effect on our results.  There may be some minor differences in the ratio of signal/foreground power after masking, however the scales where nonlinearity becomes most significant are also scales where the power spectra tend to be shot noise dominated, so we do not expect any dramatic effects from a nonlinear calculation.

\section{Conclusion}
We have presented an exploration of the effectiveness of bright pixel masking on removing foreground lines from intensity maps.  Using empirical luminosity function models and simulated intensity maps we have illustrated how masking changes the power spectra of maps for three cases: CO contaminated with HCN, $\lya$ contaminated with various atomic lines, and CII contaminated with higher order CO lines.  For the CO survey, the foreground line was faint enough on average that removing the brightest pixels significantly dropped the amplitude of the foreground spectrum.  The high angular resolution possible in the $\lya$ survey meant that the foreground contamination was limited to a few pixels which could be easily masked.  The CII survey, however, had both bright foregrounds and large pixels, so the masking was found to be ineffective.

For all of the lines, masking bright pixels altered the amplitude of the recovered power spectrum away from the desired uncontaminated value.  This means that masking loses some of the information in the spectrum.  However, in the two surveys where masking was effective, the masked spectrum had a clustering component with the same shape as the unmasked clustering spectrum.  Therefore, though the astrophysical content of the map is lost, the cosmological information contained in the shape of the clustering spectrum can be recovered from a masked map.  Thus, pixel masking seems to be a useful technique for obtaining information from even a highly contaminated CO or $\lya$ map.  If we are to obtain the remainder of the information in these surveys, it will be necessary to use some other foreground cleaning technique, such as cross correlation, or to augment it with a $P(D)$ analysis of the progressively masked power spectra.  If we are to fully unlock all of the benefits of intensity mapping surveys, it is imperative that we utilize these or other methods to isolate the signal from the foregrounds.  

The authors would like to thank Anthony Pullen, David Neufeld, and Garrett Keating for useful discussions. This work was supported by the John Templeton Foundation, the Simons Foundation, NSF grant PHY-1214000, and NASA ATP grant NNX15AB18G.

\end{document}